\documentclass[aps,twocolumn,pre,showpacs,superscriptaddress,floatfix]{revtex4}
\usepackage[dvips]{graphicx}
\usepackage{amssymb,amsmath}

\begin{document}

\title{Slow relaxation in the Ising model on a small-world network\\
with strong long-range interactions}
\author{Daun \surname{Jeong}}
\affiliation{Department of Physics, Seoul National University,
Seoul 151-747, Korea}
\author{M.Y. \surname{Choi}}
\affiliation{Department of Physics, Seoul National University,
Seoul 151-747, Korea}
\affiliation{School of Physics, Korea Institute for Advanced Study, Seoul 130-722, Korea}
\author{Hyunggyu \surname{Park}}
\affiliation{School of Physics, Korea Institute for Advanced Study, Seoul 130-722, Korea}

\begin{abstract}
We consider the Ising model on a small-world network, where the long-range interaction
strength $J_2$ is in general different from the local interaction strength $J_1$,
and examine its relaxation behaviors as well as phase transitions.
As $J_2/J_1$ is raised from zero, the critical temperature also increases, manifesting 
contributions of long-range interactions to ordering. 
However, it becomes saturated eventually at large values of $J_2/J_1$
and the system is found to display very slow relaxation, 
revealing that ordering dynamics is inhibited rather than facilitated 
by strong long-range interactions.  
To circumvent this problem, we propose a modified updating algorithm in Monte Carlo simulations,
assisting the system to reach equilibrium quickly.
\end{abstract}


\maketitle
When random links are added to a regular lattice, the latter becomes a small-world
network, characterized by short path length and high clustering~\cite{ref:network,ref:WS}.
In such a small-world network, the diameter increases very slowly with the system size:
$l\sim$ log$N$, while a regular one displays $l\sim O(N)$.
Also having common neighbors for two connected nodes is highly probable.
With these features, all elements on the network can exchange information with each other
more efficiently than on a regular lattice.
Accordingly, it is expected that dynamical systems on small-world networks may display
enhanced performance; examples include ordering in spin models~\cite{ref:Ising,ref:XY},
synchronization in coupled oscillators~\cite{ref:synch},
and computational performance of neural network~\cite{ref:neural}.

A small-world network, constructed from a one-dimensional lattice,
has two kinds of connections: (short-range) local links and (long-range) shortcuts.
It is conceivable that the two kinds of couplings in a real system have different
origins and thus different strengths; this makes it desirable to examine the general
case that interactions via local links and via shortcuts are different in strength.
It is obvious that in the absence of long-range interactions (via shortcuts),
long-range order does not emerge.  As a small amount of weak long-range
interactions is introduced, however, the system undergoes a phase transition to
the state with long-range order. This indicates the importance of shortcuts in
ordering, and it is of interest to elucidate how much they are important,
relative to local links.
On the contrary, without neighbor (local) interactions, the system cannot percolate
below a certain value of connectivity. Therefore, we conclude that high clustering due
to local interactions is also important for achieving long-range order.

To probe the roles of long- and short-range interactions in ordering,
we consider the Ising model as a prototype system exhibiting an order-disorder 
transition, and examine the transition behavior on a small-world network,
varying the long-range interaction strength $J_2$ relative to the short-range strength $J_1$.
When $J_2=0$, the system reduces to the one-dimensional Ising model and
does not display long-range order.
In the case of uniform interaction ($J_1=J_2$),
the system is known to undergo a phase transition of the mean-field
type~\cite{ref:Ising,ref:Barrat,ref:IsingOnComplex}. 
It is expected that the mean-field transition is prevalent for all finite values of
$J_2/J_1$ as long as shortcuts account for a finite fraction of total links.
Shortcuts in general assist spins to order, which is reflected by
the increase of the critical temperature; 
similar trends have been found in analytical studies of slightly different systems
in {\em equilibrium}~\cite{ref:Replica,ref:exact}. 

This work focuses on how the system approaches equilibrium, and 
reveals that strong long-range interactions ($J_2/J_1\gg1$)
gives rise to extremely slow relaxation, making
Monte Carlo (MC) dynamics based on the Metropolis algorithm inefficient.
To avoid such slow convergence, we devise a modified updating algorithm, which
assists the system to reach equilibrium more quickly.
In the limiting case that $J_1=0$, namely, all nearest-neighbor interactions are deleted,
the remaining links (shortcuts) constitute a random network with connectivity $kP$,
where $k$ denotes the range of local interaction in the underlying lattice and $P$ is
the probability of adding or re-wiring shortcuts on each local link.

Here the small-world network is constructed in the following way:
We first consider a one-dimensional (1D) lattice of $N$ nodes, each of which is
connected to its $2k$ nearest neighbors, with $k$ being the local interaction range.
Then each local edge is visited once and a random long-range connection
(shortcut) is added with probability $P$ (without removing the local edge).
Note the difference from the original Watts and Strogatz (WS)
construction~\cite{ref:WS}, where local edges are removed and reconnected to
randomly chosen nodes.

The Hamiltonian for the Ising model on a small-world network with such two kinds of
interactions is given by
\begin{equation}
H=-J_1\sum_{i}\sum_{j=1}^{k}\sigma_{i}\sigma_{i+j}
-J_2\sum_{\langle i,j \rangle}\sigma_{i}\sigma_{j},
\end{equation}
where $\sigma_i \,(=\pm 1)$ is the Ising spin on node $i$ of the network.
The first term is precisely the Hamiltonian for 1D Ising model with $k$-nearest
neighbors whereas the second one describes the contributions of spin pairs
connected via long-range connections.

We perform extensive MC simulations at various values of the addition probability $P$ 
and the coupling ratio $J_2/J_1$.  Specifically, we anneal the system, 
starting from disordered states at high temperatures, and employ 
the standard Metropolis algorithm with single-spin flip updating
to compute various quantities including the order parameter (magnetization). 
As well known, this method is expected to have the system reach efficiently the equilibrium, 
characterized by the Boltzman distribution, and to give reliable results at all temperatures 
except in the critical region, where critical slowing down is unavoidable due to strong 
fluctuations~\cite{ref:Binder}.
Obtained are results which in general support the mean-field transition and saturation
of the critical temperature, unless long-range interactions are
far stronger than local ones.

\begin{figure}
\includegraphics[width=6cm, angle=270]{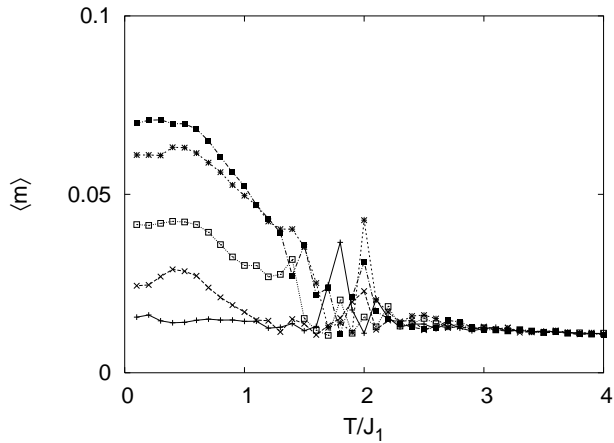}
\caption{Erratic behavior of the order parameter with temperature $T/J_1$  
(with the Boltzman constant $k_B \equiv 1$) for 
$J_2/J_1=10$ and $P=0.1$. The magnetization $m$ has been obtained from the average 
over $5\times 10^4$ MC steps after the data during initial $5\times 10^4$ MC steps discarded. 
Data, labeled by five symbols, represent the results of five MC runs, respectively, 
on a {\em single} small-world network. 
The magnetization at low temperatures varies largely run by run, 
and the system persists to remain in the disordered state.} 
\label{fig:ONR10}
\end{figure}
On the other hand, for $J_2$ much larger than $J_1$,
the order parameter (magnetization) $m$ turns out to change erratically 
around the critical temperature and the ordered phase
is hardly observed at very low temperatures.  Furthermore, at low temperatures 
it varies largely MC run by run, even though a single small-world network 
configuration is used (see Fig.~\ref{fig:ONR10} for $J_2/J_1=10$ and $P=0.1$; 
for convenience, the Boltzman constant $k_B$ is set equal to unity throughout this paper).
Namely, the result depends upon the random number sequence. 
This may be explained in the following way: At high
temperatures, all spins can flip easily and the system is in the
fully disordered state. On a small-world network, there appear
clusters which are connected by shortcuts with the interaction
strength far larger than the local (nearest neighbor) one. As the
temperature is lowered, spins on such clusters align first along
either the up- or the down-direction while other spins on the 1D
chain flip easily because thermal fluctuations are still strong
compared with local interactions. For the whole spins to be
aligned below the critical temperature, all clusters should have
the same spin orientation; otherwise some spins (which do not have
long-range interactions) may confuse between spin clusters of
different spin directions.  However, it is not probable for a spin in
the cluster to have opposite directions, due to the strong
long-range interactions at such low temperatures.  This yields low
acceptance ratios in the algorithm, resulting in extremely long
relaxation time.  Accordingly, the system tends to remain in a disordered 
state which does not correspond to the minimum of the free energy, 
even if the temperature is lower than the critical temperature.  
Finally, at very low temperatures, spins seldom flip, so that the value of
the order parameter depends on the previous history.

\begin{figure}
\includegraphics[width=6cm, angle=270]{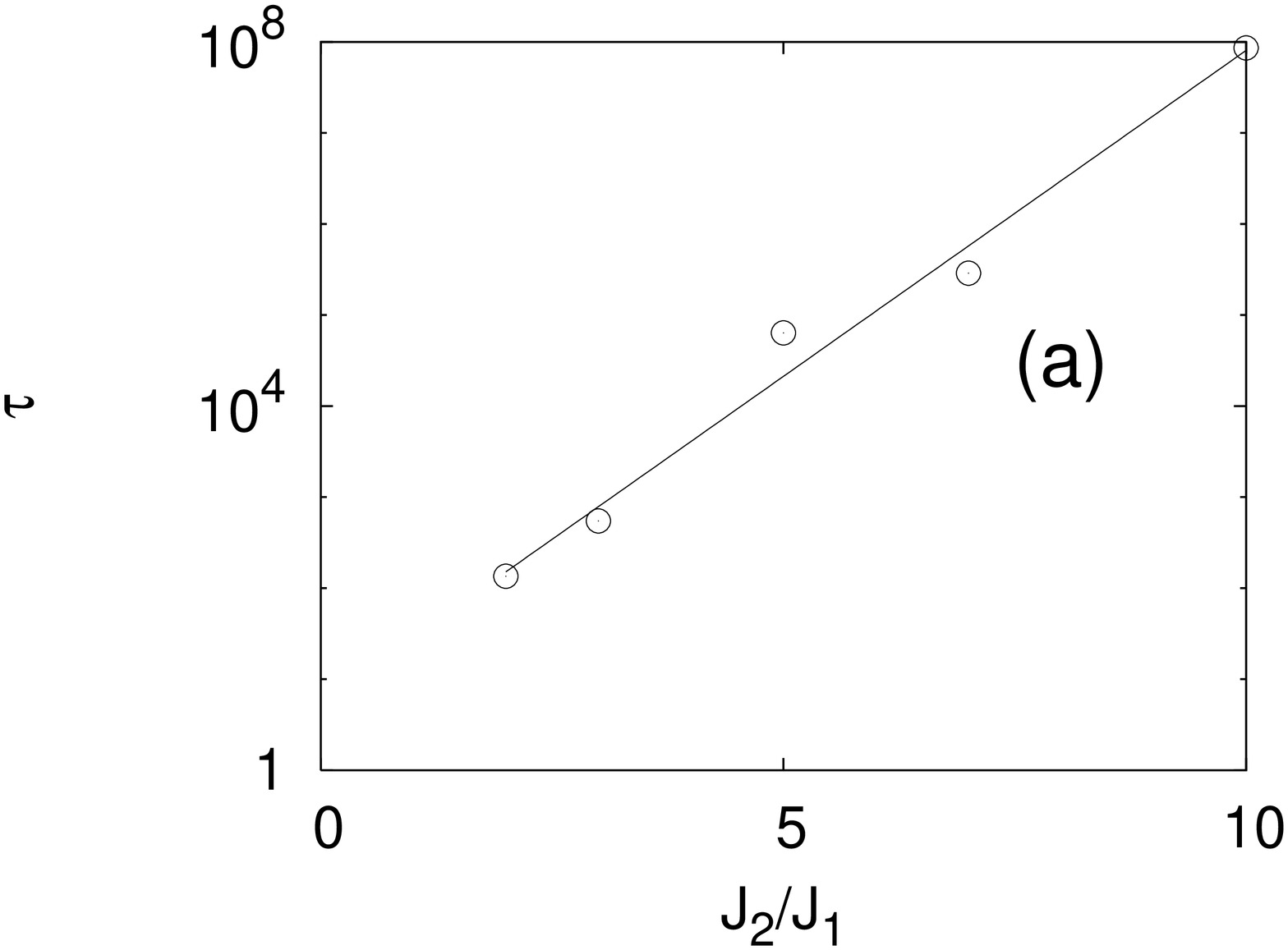}\hfill
\includegraphics[width=6cm, angle=270]{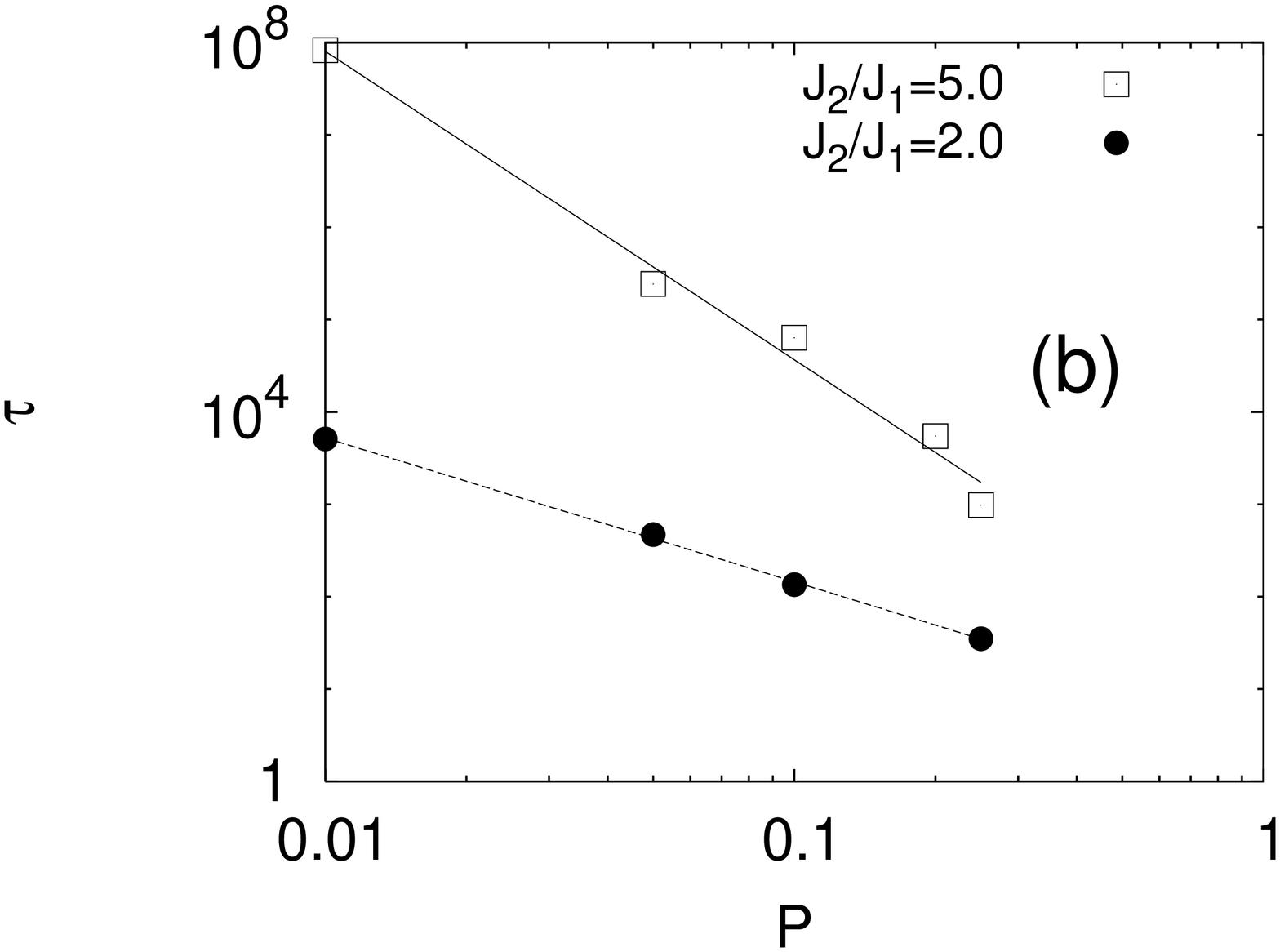}
\caption{Relaxation time $\tau$ (in units of the MC step) near the critical temperature $T_c$, 
estimated from the relation $m-m_{eq}\sim e^{-t/\tau}$, 
with $J_2/J_1$ and $P$ varied. (a) Exponential increase of 
$\tau$ with $J_2/J_1$ for $P=0.1$,
reflecting that the updating probability is an exponentially
decreasing function of $J_2/J_1$. The solid line represents the
best fit: $\tau = \tau_0 e^{aJ_2/J_1}$ with $\tau_0=1.7$ and $a =
1.6$. (b) Algebraic decrease of $\tau$ with $P$ for two values of
$J_2/J_1$, which is related with the characteristic path length of
the small-world network. The solid and dashed lines correspond to
the power-law decay $\tau = \tau_0 P^{-\sigma}$ with $\tau_0
=1.4$; $\sigma = 1.56$ and $\tau_0 = 2.8$; $\sigma = 3.3$,
respectively. } 
\label{fig:Rtime}
\end{figure}

\begin{figure}
\includegraphics[width=6cm, angle=270]{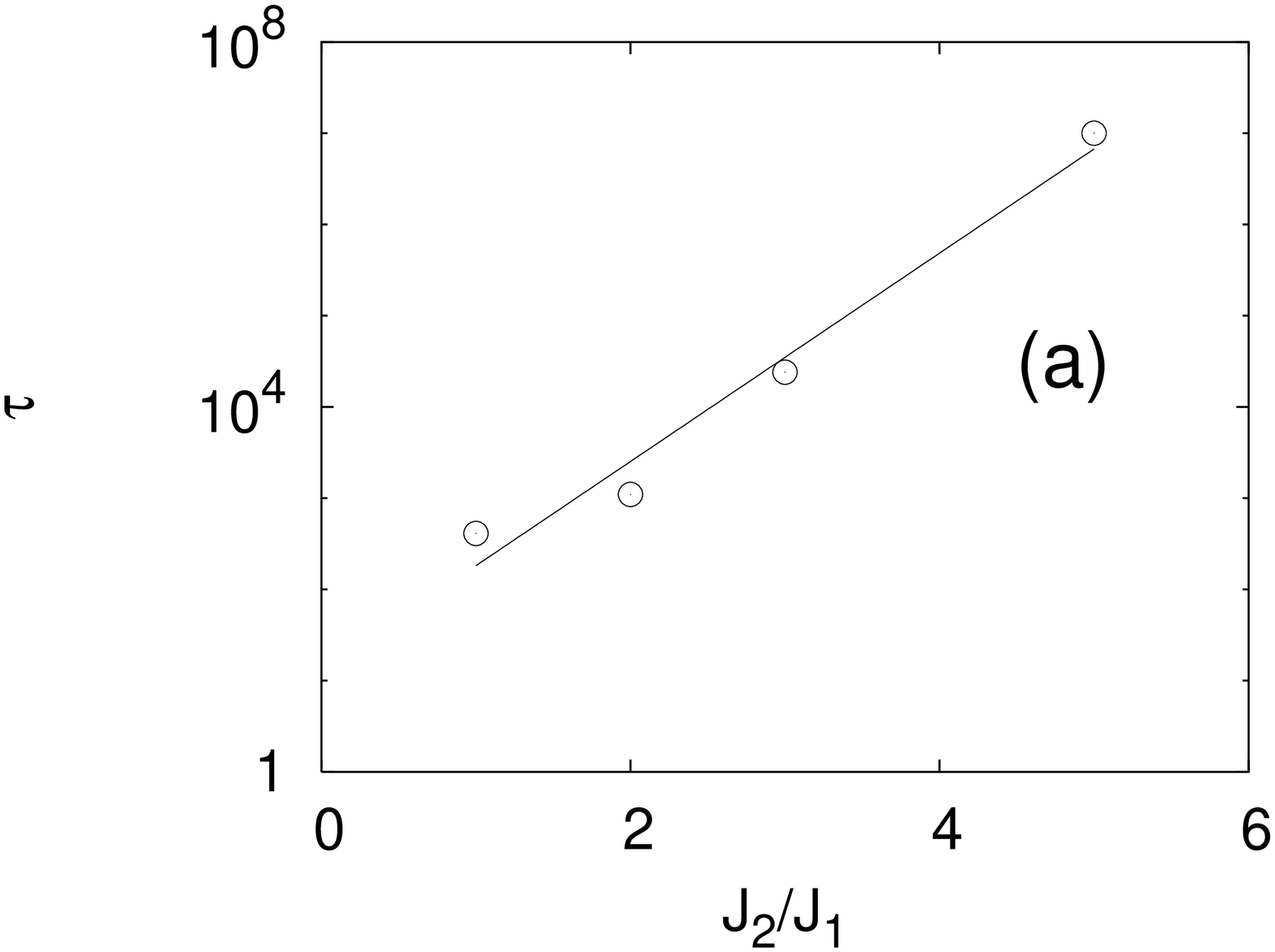}\hfill
\includegraphics[width=6cm, angle=270]{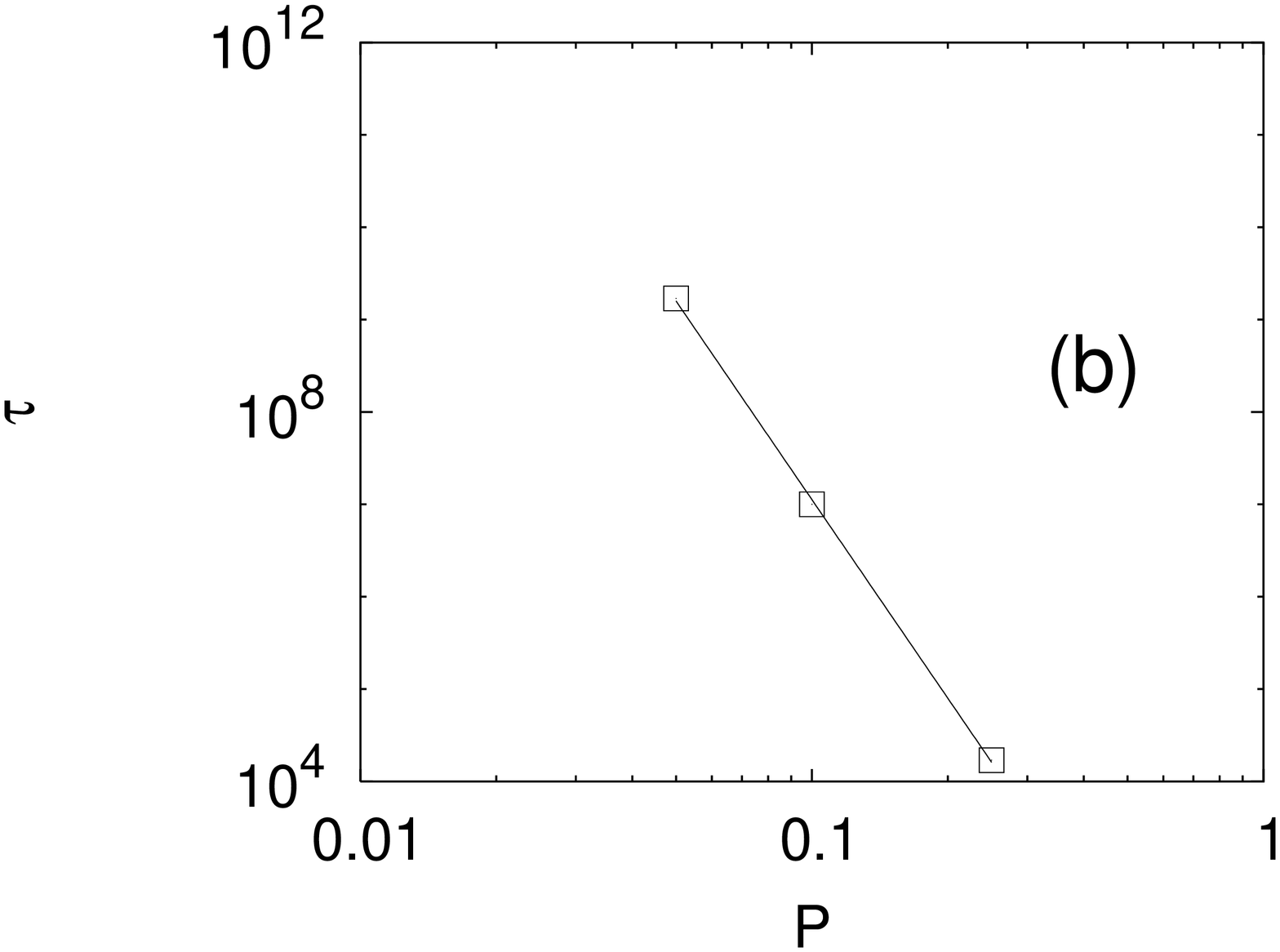}
\caption{Relaxation time $\tau$ (in units of the MC step) at low temperature $T_c/2$, 
estimated from the relation $m-m_{eq}\sim e^{-t/\tau}$, with $J_2/J_1$ 
and $P$ varied. (a) Exponential increase of $\tau$ with $J_2/J_1$ for $P=0.1$,
reflecting that the updating probability is an exponentially
decreasing function of $J_2/J_1$. The solid line represents the
best fit: $\tau = \tau_0 e^{aJ_2/J_1}$ with $\tau_0=13$ and $a=2.6$. 
(b) Algebraic decrease of $\tau$ with $P$ for $J_2/J_1 =5.0$. 
The solid line corresponds to the power-law decay 
$\tau = \tau_0 P^{-\sigma}$ with $\tau_0 =0.8$ and $\sigma = 7.1$.} 
\label{fig:Rtime2}
\end{figure}

We examine relaxation of the order parameter, starting from the fully 
ordered state near the critical temperature $T_c$ and 
from the disordered state at low temperatures, to measure the characteristic 
time scale for the system to reach equilibrium.  Assuming the exponential 
relaxation in the form $|m-m_{eq}|\sim e^{-t/\tau}$, we estimate the value 
of $\tau$, varying $J_2/J_1$ and $P$. Figure~\ref{fig:Rtime} shows the 
relaxation time $\tau$, measured in units of the MC step,
in the system of size $N=6400$, (nominally) at the 
critical temperature. 
It is observed that $\tau$ grows exponentially from $10^2$ to $10^8$ 
as $J_2/J_1$ is increased. For given value of $J_2/J_1$, $\tau$ is 
shown to depend algebraically on $P$: $\tau\sim P^{-\sigma}$.
We stress that these features are not restricted merely to the region near 
the critical temperature; they persist at all temperatures below the critical
temperature, as shown in Fig.~\ref{fig:Rtime2}. 
In fact they are {\em even more conspicuous at low temperatures};
this manifests the sharp contrast with the conventional critical slowing down,
present only near the critical temperature in systems on regular lattices~\cite{ref:Binder}. 

One can understand the exponential growth of $\tau$ in terms of the inverse 
updating probability. For large values of $J_2/J_1$, flipping one spin in a 
pair which interact strongly with each other will give much influence to
the relaxation process. 
The probability of this update is given by $e^{-\Delta E/T}$ at temperature $T$,
where the energy change $\Delta E =J_2-cJ_1$ depends on the neighboring
spin states through integer $c$. 
Since the temperature is measured in units of $J_1$, the inverse of the updating 
probability leads to the relaxation time in the form $\tau\sim e^{aJ_2/J_1}$,
where $a$ is a constant. 
On the other hand, as the link addition probability $P$ is increased, 
the characteristic path length $l$ of the system in general reduces in an
algebraic way~\cite{ref:RG}; this allows information to travel more efficiently
and thus gives rise to the algebraic decrease of $\tau$ with $P$.

Accordingly, it is concluded that the true equilibrium state may not be obtained 
within moderate MC steps when long-range interactions are substantially 
stronger than local ones. 
To circumvent this problem and to obtain the equilibrium state efficiently, 
we propose a modified updating method which is efficient in simulations of
such a system. 
The slow relaxation originates from the fact that flipping a spin interacting 
(strongly) via a shortcut is hardly probable, 
even though the free energy reduces if accepted. 
Therefore, when a spin in a cluster linked via shortcuts is selected during 
sequential update, we also consider, with probability one half, the possibility 
of flipping all the spins in the cluster simultaneously. 
Note that the probability of such cluster updating is much higher than that 
of usual single-spin updating because the energy difference involves only 
the short-range interactions. 
Still single-spin updating is also allowed, so that ergodicity of the system 
remains intact.  Further, the probability to be selected as a cluster is 
taken always the same for every relevant spin, which guarantees
the detailed balance condition. 
The new algorithm is thus expected to help the system to reach the correct 
equilibrium quickly, yielding appropriate results efficiently.

To demonstrate the efficiency of the new algorithm, we employ it to probe the case of 
strong long-range interactions ($J_2/J_1 \gtrsim 5$) 
where the conventional algorithm is practically inapplicable. 
To find the critical temperature at given values of $P$ and $J_2/J_1$, we examine
the scaling behaviors of the magnetization $m$, susceptibility $\chi$,
specific heat $C$, and Binder's cumulant~\cite{ref:Binder}.
Typically, we consider the system of size up to $N=12800$ and take the average 
over 100 different network realizations as well as the thermal average 
over $5\times 10^4$ MC steps after equilibration at each temperature. 

We write the finite-size scaling forms as
$m =N^{-\beta/\bar{\nu}}h(|t|N^{1/\bar{\nu}})$,
$\chi=N^{\gamma/\bar{\nu}}g(|t|N^{1/\bar{\nu}})$,
and $C=N^{\alpha/\bar{\nu}}f(|t|N^{1/\bar{\nu}})$
with appropriate scaling functions and critical exponents
$\gamma$, $\alpha$, $\beta$, and $\bar{\nu}$,
where $t\equiv (T-T_c)/T_c$ is the reduced temperature.
Finite-size scaling analyses of these quantities 
obtained for $N=1600, 3200, 6400$, and $12800$ unanimously
support a phase transition of the mean-field type, with exponents
$\gamma=1$, $\alpha=0$, $\beta=1/2$, and $\bar{\nu}=2$. The
critical temperature turns out to agree well with the value
obtained from the unique crossing point of Binder's cumulant.
It is thus concluded that the system undergoes a finite-temperature 
transition of mean-field nature for $J_2/J_1>0$ and $P\neq0$.

\begin{figure}
\includegraphics[width=7cm, angle=270]{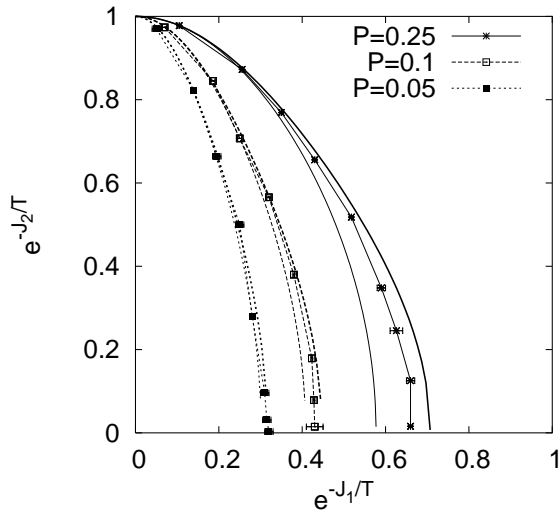}
\caption{Phase diagram of the Ising model on a small-world
network, where the region below each boundary represents the
ordered phase for the corresponding value of the addition
probability $P$. Simulation data for various values of $P$ are
depicted by symbols on lines; the latter are merely guides to
eyes.
Analytic results in Refs.~\onlinecite{ref:Replica} and~\onlinecite{ref:exact} 
are also plotted, with the same kinds of thick and thin lines, respectively,
for each value of $P$. 
They coincide with numerical results when $P$ is small and/or $J_2$ is sufficiently 
smaller than $J_1$. 
For $J_2/J_1$ large, our data locate between the two analytic results 
in the phase diagram.} 
\label{fig:k=1}
\end{figure}

Here shortcut interactions are essential for the 1D system to display long-range order. 
The critical temperature $T_c/J_1$ is expected to increase as $J_2/J_1$ is raised. 
In simulations, however, $T_c/J_1$ does not keep
increasing with $J_2/J_1$ beyond a certain value depending on $P$.
In Fig.~\ref{fig:k=1}, we present the phase diagram of the system with range $k=1$, 
for various values of $P$. 
In this case of $k=1$, analytic results have been reported for similar systems:
A replica symmetric solution has been developed on the networks 
constructed by superimposing random graphs onto a one-dimensional ring~\cite{ref:Replica}. 
Subsequently, combinatorics has been used to treat quenched disorder on the networks, 
where each node is restricted from having more than one shortcut~\cite{ref:exact}. 
Those networks coincide with our network only in the limit $P \rightarrow 0$. 
For finite $P$, in contrast to the latter, 
we allow each node to have more than one shortcut in the construction, 
which is more realistic and necessary for the small-world 
network to have an exponential tail in the degree distribution. 
Further, one end of each added shortcut is determined sequentially, 
which makes our network have less numbers of large-degree nodes than 
the former (superimposed random) network. 
Accordingly, the standard small-world network used in this study lies in between the two 
types of network in Refs.~\onlinecite{ref:Replica} and~\onlinecite{ref:exact}.
Since spins on those nodes which have more links facilitate more spins to order,
the critical temperature of the system on the small-world network should be 
lower than that in Ref.~\onlinecite{ref:exact} and higher that that 
in Ref.~\onlinecite{ref:Replica}, 
and such difference is expected to grow as $P$ and $J_2$ are increased.
It is indeed observed in Fig.~\ref{fig:k=1} that the phase boundary of the system 
on the small-world network locates in between the boundary obtained 
in Ref.~\onlinecite{ref:exact} and that in Ref.~\onlinecite{ref:Replica}, 
particularly in case that $J_2$ is substantially larger than $J_1$ 
and $P$ is not very small ($P \gtrsim 0.05$). 
\begin{figure}
\includegraphics[width=7cm, angle=270]{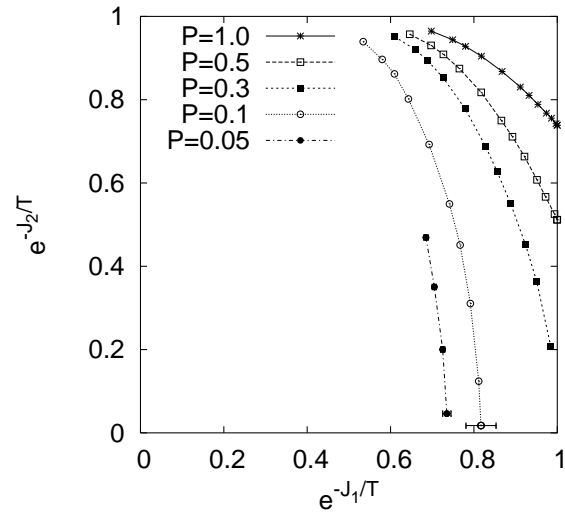}
\caption{Phase diagram of the Ising model on a small-world network
with range $k=2$. Simulation data for various values of $P$ are
depicted by symbols on lines; the latter are merely guides to
eyes. For $P\geq 0.3$, the phase boundary intersects the $J_1=0$ line 
at a finite value of $J_2$, manifesting the presence of a phase transition. 
This exhibits that a small-world network with local links deleted has 
a threshold value of $P$ below which no long-range order emerges.}
\label{fig:k=2}
\end{figure}

We also consider the system with range $k=2$, where local interactions are
present between the next nearest neighbors as well as the nearest neighbors,
and perform extensive simulations, the results of which are displayed in 
Fig.~\ref{fig:k=2}. 
As expected, the region of the ordered phase in the phase diagram is increased 
compared with the case $k=1$. 
Except for this, when $P$ is small ($P<0.3$), the overall features are entirely 
similar to those of the case $k=1$: The critical temperature increases with 
$J_2/J_1$, eventually saturating to a finite value. 
In case that $P\geq 0.3$, on the other hand, one observes an order-disorder transition
on the $J_1=0$ line; this corresponds to the small-world network whose local links
are all deleted so that there remain only randomly added shortcuts with fraction $P$. 
In comparison with the case $P<0.3$, where no ordered phase exists on this line, 
manifested is the percolation problem in the resulting random graph. 
Namely, the system is percolating only when its connectivity, given by $kP$, 
is higher than $2P_c \approx 0.6$. It is pleasing that this value agrees with the 
known expression for the threshold value $P_c=1-\sqrt{(k-1)/k}$~\cite{ref:Barrat}. 
We have also performed simulations of the system with $k=3$, to obtain fully
consistent results.
It is of interest that the threshold value is smaller than that of the Erd\"os-Renyi
(ER) random graph~\cite{ref:ER}, which reflects that our random graph is still 
more regular than the ER graph.

In summary, we have studied via extensive numerical simulations the Ising model 
on a small-world network, where long-range interactions via shortcuts are 
in general different from local interactions. 
It has been demonstrated that long-range interactions via added shortcuts 
help spins to order, raising the critical temperature at first and having it 
saturated eventually. 
Of particular interest is the case of strong long-range interactions,
relative to the local ones, where each cluster may play the role of temporarily 
quenched randomness. The system then tends to be trapped in a local minimum,
inhibited from relaxation to the global minimum (i.e., equilibrium);
this results in very slow relaxation, making simulations inefficient. 
This is in contrast with the Ising model on conventional regular or disordered lattices, 
where severe inhomogeneity in the interaction strength is absent and  
equilibrium is reached quickly at all temperatures except in the critical region
without any erratic behavior. 
To circumvent this problem, we have developed a modified updating algorithm, 
assisting the system to reach equilibrium quickly.
Any dynamical system on a small-world network with strong long-range
interactions is expected to behave similarly,
and the modified algorithm developed here may be used to
obtain (equilibrium) thermodynamic properties efficiently.
Finally, it would be of interest to investigate the case that 
long- and short-range interactions have opposite signs ($J_2/J_1 < 0$). 
The coexistence of ferromagnetic and antiferromagnetic interactions in general 
introduces frustration into the system, which, together with the randomness associated with
the long-range connections, may lead to the (truly) glassy behavior~\cite{comm}.
The detailed investigation of how such a glass system relaxes depending
on the value $J_2 /J_1$ and comparison with the other cases
are left for further study.

\acknowledgments
This work was supported in part by the KOSEF Grant No. R01-2002-000-00285-0 
and by the BK21 Program.


\begin{thebibliography}{10}

\bibitem{ref:network}For reviews, see, e.g.,
Science {\bf 284}, 79-109 (1999);
M.E.J. Newman, J. Stat. Phys. {\bf 101},  819  (2000);
D.J. Watts, {\em Small Worlds} (Princeton University Press, Princeton, 1999);
S.H. Strogatz, Nature {\bf 410},  268  (2001);
R. Albert and A.-L. Barab{\'a}si, Rev. Mod. Phys. {\bf 74}, 47 (2002),
S.N. Dorogovtsev and J.F.F. Mendes, Adv. Phys. {\bf 51}, 1079 (2002).

\bibitem{ref:WS} D.J. Watts and S.H. Strogatz, Nature {\bf 393}, 440 (1998).

\bibitem{ref:Ising}
M. Gitterman, J. Phys. A: Math. Gen. {\bf 33}, 8373 (2000);
A. P\c{e}kalski, Phys. Rev. E {\bf 64}, 057104 (2001);
H. Hong, B.J. Kim, M.Y. Choi, {\it ibid.} {\bf 66}, 018101 (2002).

\bibitem{ref:XY}
B.J. Kim, H. Hong, P. Holme, G.S. Jeon, P. Minnhagen, M.Y. Choi,
Phys. Rev. E {\bf 64}, 056135 (2001).

\bibitem{ref:synch}
H. Hong, M.Y. Choi, and B.J. Kim, Phys. Rev. E {\bf 65}, 026139 (2002).

\bibitem{ref:neural}
L.G. Morelli, G. Abramson, and M.N. Kuperman, Eur. Phys. J. B {\bf 38}, 495 (2004);
P.N. McGraw and M. Menzinger, Phys. Rev. E {\bf 68}, 047102 (2003).

\bibitem{ref:Barrat}
A. Barrat and M. Weigt, Eur. Phys. J. B {\bf 13}, 547 (2000).

\bibitem{ref:IsingOnComplex}
The Ising model has also been examined analytically and numerically 
on other types of complex network.  See
S.N. Dorogovtsev, A.V. Goltsev, and J.F.F. Mendes, Phys. Rev. E {\bf 66}, 016104 (2002);
M. Leone, A. V{\'a}zquez, A. Vespignani, and R. Zecchina, Eur. Phys. J. B {\bf 28}, 191 (2002);
A. Ramezanpour, Phys. Rev. E {\bf 69}, 066114 (2004).

\bibitem{ref:Replica}
T. Nikoletopoulos, A.C.C. Coolen, I. P\'erez Castillo, N.S. Skantzos, J.P.L. Hatchett, 
and B. Wemmenhove, J. Phys. A: Math. Gen. {\bf 37}, 6455 (2004). 

\bibitem{ref:exact}
J.V. Lopes, Y.G. Pogorelov, J.M.B. Lopes dos Santos, and R. Toral, 
Phys. Rev. E {\bf 70}, 026112 (2004).

\bibitem{ref:Binder}
See, e.g., K. Binder (eds.), {\it Applications of the Monte Carlo Method in Statistical Physics} 
(Springer-Verlag, Berlin, 1987); 
K. Binder and D.W. Heermann,
{\it Monte Carlo Simulation in Statistical Physics} (Springer-Verlag, Berlin, 1992).

\bibitem{ref:RG}
M.E.J. Newman, D.J. Watts, Phys. Lett. A {\bf 263}, 341 (1999).

\bibitem{ref:ER}
B. Bollob\'as, {\em Random Graphs} (Academic Press, London, 1985).

\bibitem{comm}
Note the difference from the usual case that the long-range interaction $J_2$ itself 
is randomly distributed, taking to values $\pm J$; the resulting spin-glass phase 
has been examined {\em in equilibrium}.  See Ref.~\onlinecite{ref:Replica}. 

\end{thebibliography}
\end{document}